\documentclass[journal,draftcls,onecolumn,12pt,twoside]{IEEEtranTCOM}
\normalsize

\usepackage{cite}

\usepackage{graphicx}
\DeclareGraphicsExtensions{.pdf,.jpeg,.png,.jpg, .svg}

\usepackage{amssymb}
\usepackage[cmex10]{amsmath}
\usepackage{array}

\usepackage[font = small]{caption}
\usepackage{setspace}
\usepackage{color}
\usepackage{commath}
\usepackage{amssymb}
\usepackage{bm}
\usepackage{amsmath}

\usepackage{mathtools}

\newcommand{\etal}{\textit{et al}. }
\newcommand{\ie}{\textit{i}.\textit{e}. }

\usepackage[utf8]{inputenc}
\usepackage{algorithm}

\usepackage[]{algpseudocode}

\algnewcommand\algorithmicinput{\textbf{Input:}}
\algnewcommand\Input{\item[\algorithmicinput]}
\algnewcommand\algorithmicoutput{\textbf{Output:}}
\algnewcommand\Output{\item[\algorithmicoutput]}
\algnewcommand\algorithmichline{}
\algnewcommand\Hline{\item[\algorithmichline]}
\makeatletter
    \newcommand*{\algrule}[1][\algorithmicindent]{\makebox[#1][l]{\hspace*{.5em}\thealgruleextra\vrule height \thealgruleheight depth \thealgruledepth}}%
\newcommand*{\thealgruleextra}{}
\newcommand*{\thealgruleheight}{.85\baselineskip}
\newcommand*{\thealgruledepth}{.25\baselineskip}

\newcount\ALG@printindent@tempcnta
\def\ALG@printindent{%
    \ifnum \theALG@nested>0
        \ifx\ALG@text\ALG@x@notext
        \else
            \unskip
            \addvspace{-1pt}
            \ALG@printindent@tempcnta=1
            \loop
                \algrule[\csname ALG@ind@\the\ALG@printindent@tempcnta\endcsname]%
                \advance \ALG@printindent@tempcnta 1
            \ifnum \ALG@printindent@tempcnta<\numexpr\theALG@nested+1\relax
            \repeat
        \fi
    \fi
    }%
\usepackage{etoolbox}
\patchcmd{\ALG@doentity}{\noindent\hskip\ALG@tlm}{\ALG@printindent}{}{\errmessage{failed to patch}}
\makeatother

\newbox\statebox
\newcommand{\myState}[1]{%
    \setbox\statebox=\vbox{#1}%
    \edef\thealgruleheight{\dimexpr \the\ht\statebox+1pt\relax}%
    \edef\thealgruledepth{\dimexpr \the\dp\statebox+1pt\relax}%
    \ifdim\thealgruleheight<.75\baselineskip
        \def\thealgruleheight{\dimexpr .75\baselineskip+1pt\relax}%
    \fi
    \ifdim\thealgruledepth<.25\baselineskip
        \def\thealgruledepth{\dimexpr .25\baselineskip+1pt\relax}%
    \fi
    \State #1%
    \def\thealgruleheight{\dimexpr .75\baselineskip+1pt\relax}%
    \def\thealgruledepth{\dimexpr .25\baselineskip+1pt\relax}%
}

\algdef{SE}[DOWHILE]{Do}{doWhile}{\algorithmicdo}[1]{\algorithmicwhile\ #1}%

\begin{document}
\title{Energy-Efficient Resource Allocation for Elastic Optical Networks using Convex Optimization}
\author{Mohammad~Hadi,~\IEEEmembership{Member,~IEEE,}
        and~Mohammad~Reza~Pakravan,~\IEEEmembership{Member,~IEEE}
}

\maketitle

\begin{abstract}
We propose a two-stage algorithm for energy-efficient resource allocation constrained to QoS and physical requirements in OFDM-based EONs. The first stage deals with routing, grooming and traffic ordering and aims at minimizing amplifier power consumption and number of active transponders. We provide a heuristic procedure which yields an acceptable solution for the complex ILP formulation of the routing and grooming. In the second stage, we optimize transponder configuration including spectrum and transmit power parameters to minimize transponder power consumption. We show how QoS and transponder power consumption are represented by convex expressions and use the results to formulate a convex problem for configuring transponders in which transmit optical power is an optimization variable. Simulation results demonstrate that the power consumption is reduced by $9\%$ when the proposed routing and grooming algorithm is applied to European Cost239 network with aggregate traffic $60$ Tbps. It is shown that our convex formulation for transponder parameter assignment is considerably faster than its MINLP counterpart and its ability to optimize transmit optical power improves transponder power consumption by $8\%$ for aggregate traffic $60$ Tbps. Furthermore, we investigate the effect of adaptive modulation assignment and transponder capacity on inherent tradeoff between network CAPEX and OPEX. 
\end{abstract}


\IEEEpeerreviewmaketitle

\section{Introduction}\label{sec_I}
\IEEEPARstart{E}{nergy} consumption of equipments in telecommunication networks has been recognized as one of major contributors in overall future energy consumption. Traditional fixed grid optical networks are not adaptive and therefore, they can not efficiently use system resources such as transmit optical power and spectrum to minimize network energy consumption. Elastic Optical Networks (EON) can provide an energy-efficient network configuration by adapting the allocated resources to the communication demands and physical conditions. It has been shown that the cost and energy savings of migration to an energy-efficient EON outweigh the added cost of the equipments \cite{palkopoulou2012quantifying}. Higher energy efficiency of Orthogonal Frequency Division Multiplex (OFDM)-based optical networks has been demonstrated in \cite{nag2011spectrum} while the outstanding features such as scalability, compatibaility and adaptation which nominate OFDM as the enabling technology of EONs have been reported in \cite{armstrong2009ofdm}.

Many variants of optimization formulations and algorithms have been proposed for resource assignment in optical networks \cite{chatterjee2015routing, abkenar2017study, sambo2015routing, hadi2016improved}. Optimized Routing and Wavelength Assignment (RWA) in nonlinear Wavelength Division Multiplex (WDM) networks has been studied in \cite{roberts2016convex, ives2014physical, ives2014adapting, ives2015routing} which is not applicable to EON with variable assignment of spectrum width and optical carrier frequencies. Some of the proposed solutions for resource allocation in EONs do not consider the Quality of Service (QoS) requirements in their optimization analysis \cite{fallahpour2015energy, hadi2016improved, archambault2016routing} which is a necessary parameter for practical network planning. Some of the researchers have proposed a general nonlinear expression to include QoS requirement as a fundamental constraint of the resource assignment optimization problem \cite{zhao2015nonlinear, yan2015resource, yan2017joint, hadi2017resource} but several of them have focused on energy-efficiency \cite{khodakarami2016quality, khodakarami2014flexible}. Khodakarami \etal have proposed a Mixed-Integer NonLinear Program (MINLP) for energy-efficient resource allocation in EONs which is decomposed into three sub-problems with heuristic solutions \cite{khodakarami2014flexible}. Their general formulation selects the best configuration for transponders and grooming switches such that the total network power consumption is minimized. Although they consider QoS requirement, they do not take into account the transmit optical power as an optimization variable. Traffic grooming is employed to reuse active physical components in the network and thereby increase the energy efficiency \cite{hasan2013study, khodakarami2014flexible}. However, there is no comprehensive work that simultaneously considers traffic grooming along with adaptive assignment of spectrum and transmit power variables and investigates their effect on different network elements.

A convex optimization is a type of mathematical optimization problems characterized by convex objective and constraint functions. Because of their desirable properties, convex optimization problems can be solved with a variety of efficient algorithms. There are software packages that can be used to solve large scale convex optimization problems \cite{boyd2007tutorial, boyd2004convex}. Using Geometric Programming (GP) convexification techniques, communication optimization problems can be converted to convex form to allow us to use fast and efficient convex optimizer software packages \cite{hadi2017resource, chiang2005geometric}. 

The introduction of OFDM-based EONs, as a promising solution for developing green core optical networks, has raised new networking challenges and increased resource provisioning complexity. This mandates designing efficient and fast algorithms to utilize network adaptation capabilities for reducing power consumption. In this paper, we consider the energy-efficient resource allocation in OFDM-based EONs and following the approach of \cite{yan2015resource, khodakarami2014flexible}, we decompose it into two interconnected optimization sub-problems 1) Routing, Grooming and Ordering (RGO) 2) Transponder Parameter Assignment (TPA). RGO is responsible for routing the traffic along the best possible path while minimizing the number of assigned lightpaths and used transponders by aggregating traffic partitions into a traffic segment closer to the information bit rate of the optical transponders. It also defines the orders of traffic routes on each link. In TPA, transponder variables such as transmit optical power, modulation level, coding rate, number of sub-carriers and location of optical carrier frequency are optimally configured such that total transponder power consumption is minimized, QoS requirements are guaranteed and physical constraints are satisfied. We formulate RGO as an Integer Linear Program (ILP) and provide a heuristic approach which yields its near-optimum solution in a reasonable time. Long and high volume traffic requests are more susceptible to transmission impairments and have considerable power consumption and spectrum usage. In RGO, we mainly focus on such long and high volume traffic requests and groom them over shorter lightpaths to reduce their high amount of power consumption and spectrum usage. TPA is conventionally formulated as a complex MINLP  which is an NP-hard problem. A key contribution of this paper is that we have proposed a method to covert the complex MINLP to a convex problem which can be solved with readily available convex optimization algorithms \cite{boyd2004convex}. In the proposed convex formulation, joint optimization of power and spectrum variables are addressed. We consider Optical Signal to Noise Ratio (OSNR) as an indicator of QoS and show how the required levels of QoS can be expressed by a convex inequality constraint. We also provide a convex expression for power consumption of transponders. Simulation results show that the proposed convex formulation provides the same results as its MINLP counterpart but in a much shorter time. Simulation results also demonstrate that optimizing transmit optical power considerably improves the power consumption of different network elements compared to the fix assignment of the transmit optical power. As another result, we demonstrate that it is not required to have a complex transponder supporting high number of different modulation levels and coding rates. We also show that how transponder capacity affects the expected power saving of our proposed algorithm for RGO. Furthermore, we reveal the inherent tradeoff between CAPital EXpenditure (CAPEX) and OPerational EXpenditure (OPEX) as the transponder capacity increases and show that increasing transponder capacity increases CAPEX while results in lower amount of OPEX.

The rest of the paper is organized as follows. System model is introduced in Section \ref{Sec_II}. In Sections \ref{Sec_III}, we propose our two-stage resource allocation algorithm. Section \ref{Sec_IV} is devoted to RGO sub-problem while in Section \ref{Sec_V}, we show how TPA sub-problem is formulated as a convex optimization problem. Simulation results are included in \ref{Sec_VI}. Finally, we conclude the paper in Section \ref{Sec_VII}.

\textbf{Notation}. Optimization variables are shown in lower case letters. Bold letters are used to denote vectors and sets. We use calligraphic font to show constants. Functions are also shown by capital letters and their arguments are given by lower indexes. Optical fiber characterizing parameters are shown by their typical notations.
\section{System Model}\label{Sec_II}
Consider an optical network and assume that $\mathbf{V}$ and $\mathbf{L}$ denote its sets of optical nodes and directional optical fiber links. Optical fiber $l$ begins at node $B(l)$, ends at node $E(l)$ and has length $\mathcal{L}_l$. The optical fiber bandwidth $\mathcal{B}$ is assumed to be gridless. $\mathbf{Q}$ is the set of traffic requests. Each traffic request originates from source node $S(q)$, terminates at destination node  $D(q)$ and has volume of  $\mathcal{R}_q$ bit per second. As shown in Fig. \ref{fig:groom}, each optical node has a bank of transmit and receive transponders and an electronic grooming switch. Input/output traffic requests and added/dropped traffics can be merged in grooming switch to reduce the unused capacity of transponders and number of active transponders\cite{khodakarami2014flexible}. There are at most $\mathcal{T}$ pairs of transmit and receive transponders in each bank which are indexed by $\mathbf{T}=\{1,\cdots,\mathcal{T}\}$.  $f_{(v,t,q)}$ shows the part of $q$th traffic request that may be added at $t$th transmit transponder of node $v$, \ie transmit transponder $(v,t)$. Similarly, $c_{(v,t,q)}$ denotes the part of $q$th traffic request that may be dropped at receive transponder $(v,t)$. $x_{(v,t,l)}$ is a binary variable that shows if link $l$ is used for routing the traffics added by transmit transponder $(v,t)$. Furthermore, binary variable $d_{(v,t,v',t')}$ shows that if the added traffic of transmit transponder $(v,t)$ is dropped at receive transponder $(v',t')$. Each transmit transponder is given a contiguous spectrum bandwidth $\Delta_{(v,t)}$ around optical carrier frequency $\omega_{(v,t)}$. The assigned contiguous spectrum bandwidth includes $2^{b_{(v,t)}}$ OFDM sub-carriers with sub-carrier space of $\mathcal{F}$ so, $\Delta_{(v,t)}=2^{b_{(v,t)}}\mathcal{F}$. To facilitate optical switching and remove the high cost of spectrum conversion \cite{spectrum2017hadi}, we assume that the assigned spectrum bandwidth to each transponder is continuous over its routed path. The added traffic of transponder $(v,t)$ passes $\mathcal{N}_{(v,t)}$ equal-length fiber spans along its routed path and has $\mathcal{N}_{(v,t, v',t')}$ shared fiber spans with transmit transponder $(v',t') \neq (v,t)$.  Each fiber span has fixed length $\mathcal{L}_{spn}$ and is equipped with its own optical amplifier to compensate for the attenuation. $\mathbf{H}_l$ shows the set of transmit transponders that share optical fiber $l$ on their allocated routes. There are pre-defined modulation levels $c$ and coding rates $r$ where each pair of $(c, r)$ requires minimum OSNR $\Theta(c, r)$ to get a pre-Forward-Error-Correction (FEC) Bit-Error-Rate (BER) value of $1\times 10^{-4}$, as shown in Tab. \ref{tab:snr_spc} \cite{yan2015resource}. The optical transponder $(v,t)$ is given a modulation level $c_{(v,t)}$ and a coding rate $r_{(v,t)}$ and fills its assigned optical bandwidth $\Delta_{(v,t)}$ with optical power $p_{(v,t)}$. Transponders have maximum information bit rate $\mathcal{C}$ and fill both of the polarizations with the same power. The configuration of each receive transponder is the same as its corresponding transmit transponder. According to filtering and switching constraints in optical nodes, we consider a guard band $\mathcal{G}$ between any two adjacent spectrum bandwidths of a link. 
\begin{table}[t!]
\small
\centering
\caption{Available modulation levels and coding rates along with their corresponding minimum required OSNR $\Theta(c,r)$ to achieve a pre-FEC BER values of $1 \times 10^{-4}$.}\label{tab:snr_spc}
\begin{tabular}{ccccccc}
\hline
\hline
$r$/$c$ & $1$& $2$ & $3$ & $4$ & $5$ & $6$  \\
\hline
$2/3$ & $1.5$& $2.3$ & $5.9$ & $9.1$ & $17.4$ & $28.8$  \\
$3/4$ & $1.7$& $2.9$ & $7.8$ & $12.0$ & $24.0$ & $40.7$  \\
$8/9$ & $3.6$& $4.6$ & $12.9$ & $20.9$ & $42.7$ & $75.8$  \\
\hline
\hline
\end{tabular}
\end{table}
\begin{figure}[t!] 
\center{\includegraphics[scale=0.55]{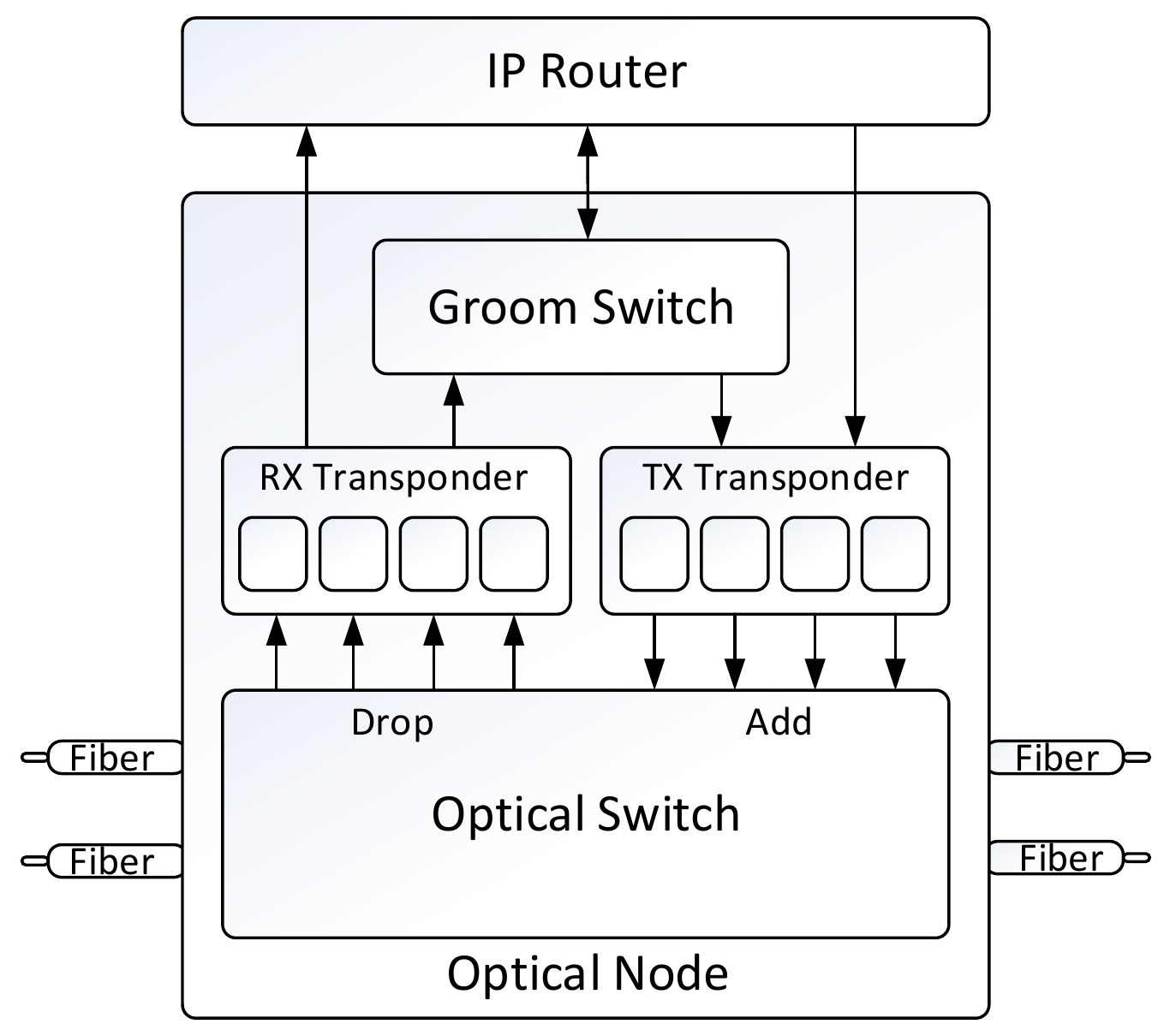}}
\center{\caption{\label{fig:transponder} Optical node block diagram which has a bank of transmit and receive transponders and an electronic grooming switch.}}
\end{figure}

Transponders, grooming switches and optical amplifiers are the main sources of network power consumption \cite{khodakarami2014flexible, fallahpour2015energy} so, total network power consumption is given by:
\begin{equation}\label{eq:tot_pow}
\sum\limits_{v,t}X_{(v,t)}+\sum\limits_{v}G_{(v)}+\sum\limits_{l}A_{(l)}
\end{equation}
where $X_{(v,t)}$ is total power consumption of each linked pair of transmit and receive transponders, $G_{(v)}$ is power consumption of grooming switch $v$ and $A_l$ shows the total power consumption of optical amplifiers over link $l$. Considering the architecture of Fig. \ref{fig:transponder}, $X_{(v,t)}$ can be calculated as follows:
\begin{align}\label{eq:trx_pow}
X_{(v,t)}= \mathcal{P}_{txb}+\frac{\mathcal{P}_{enc}}{r_{(v,t)}}+\frac{\mathcal{P}_{fft}}{2}2^{b_{(v,t)}}b_{(v,t)}+\mathcal{P}_{rxb}+\frac{\mathcal{P}_{dec}}{r_{(v,t)}}+\frac{\mathcal{P}_{fft}}{2}2^{b_{(v,t)}}b_{(v,t)}+2^{b_{(v,t)}} \mathcal{P}_{dsp}
\end{align}
where $\mathcal{P}_{txb}$ and $\mathcal{P}_{rxb}$ are transmit and receive transponder bias terms (which includes the power consumption of bit interleaver, laser, A/D converter, D/A converter, Mach–Zehnder modulator, serial/parallel, parallel/serial), $\mathcal{P}_{enc}$ and $\mathcal{P}_{dec}$ are the scaling coefficient of encoder and decoder power consumptions, $\mathcal{P}_{fft}$ denotes the power consumption for a two point FFT operation and $\mathcal{P}_{dsp}$ is the power consumption scaling coefficient of the receiver DSP operations. The electronic grooming switch consumes $\mathcal{E}_{grm}$ per each switched bit \cite{khodakarami2014flexible}. Considering the total groomed traffic at node $v$, we can write \cite{khodakarami2014flexible}:
\begin{align}\label{eq:grm_pow}
G_{(v)}=\mathcal{E}_{grm}\sum\limits_{t,q}[f_{(v,t,q)}+c_{(v,t,q)}]
 -\mathcal{E}_{grm}\sum\limits_{q}R_q[\delta(v-S(q))+\delta(v-D(q))]
\end{align}
where $\delta()$ is discrete impulse function. Each optical amplifier also needs $\mathcal{P}_{amp}$ input power, therefore the total power consumption of optical amplifiers over an active link is \cite{khodakarami2014flexible}:
\begin{equation}\label{eq:amp_pow}
A_{(l)}=\mathcal{P}_{amp}\Big[\frac{\mathcal{L}_l}{\mathcal{L}_{spn}}+1\Big]\Big[1-\delta(\sum\limits_{v,t}x_{(v,t,l)})\Big]
\end{equation}
\begin{figure}[t!] 
\center{\includegraphics[scale=0.5]{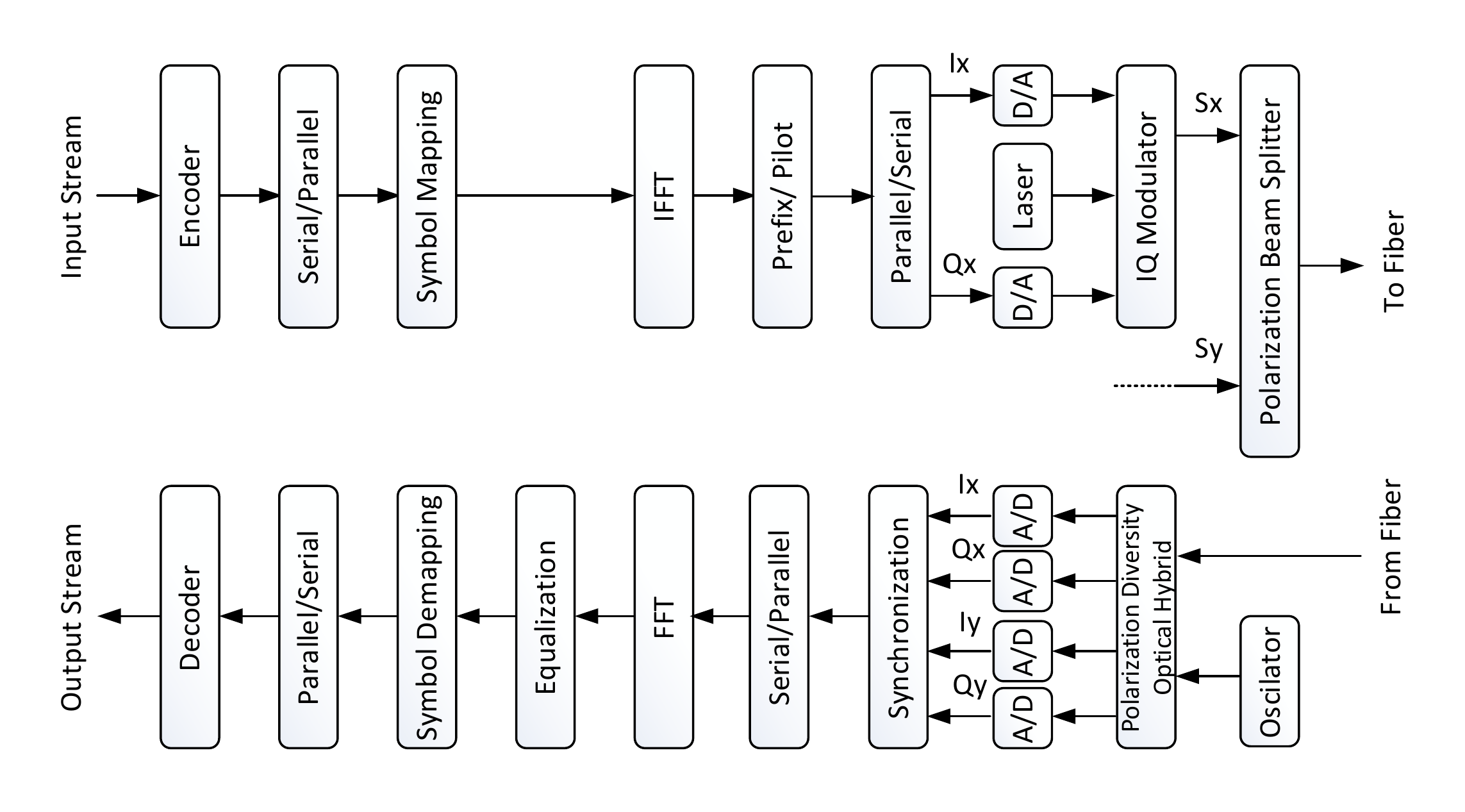}}
\center{\caption{\label{fig:transponder} Block diagram of a pair of OFDM-based transmit and receive transponders.}}
\end{figure}
\section{Resource Allocation Problem}\label{Sec_III}
To have a green EON, we need a resource allocation algorithm to determine the values of system model variables such that the network consumes the minimum power while physical constrains are satisfied and desired levels of QoS (or equivalently desired levels of OSNR) are guaranteed. In general, such a problem is modeled as an optimization problem and it is shown to be NP-hard \cite{yan2015resource}. Therefore, we propose a two-stage algorithm in which the complex resource allocation problem is decomposed into two sub-problems which are named RGO and TPA. The pseudo code of this algorithm is shown in Alg. \ref{alg:rsa}. Solving RGO sub-problem, the connection requests are routed, the grooming switches are configured and the order of traffics on links are determined. Transponder parameters such as transmit optical power, modulation level, coding rate, number of sub-carriers and optical carrier frequency are optimized during TPA sub-problem. The power consumption of optical amplifiers and grooming switches are respectively related to the traffic routes and configuration of the grooming switches which are mainly addressed in RGO sub-problem. On the other hand, total transponder power consumption are minimized by solving TPA sub-problem. Usually the search for a near optimal solution involves iterations between these two sub-problems. To save this iteration time, it is of great interest to hold the run time of each sub-problem at its minimum value. 

In the coming sections, we formulate RGO problem as an ILP and then provide a heuristic solution for it. We also introduce an MINLP formulation for TPA sub-problem and show how this complex formulation can be converted to a convex optimization problem.
\begin{algorithm}[!htbp]
\setstretch{0.9}
\small
\caption{Resource Allocation Algorithm}\label{alg:rsa}
\begin{algorithmic}[1]
\Input{network topology, traffic matrix, link parameters, noise and nonlinearity parameters, transponder parameters}
\Output{routes, groomed traffics, number of subcarriers, optical carrier frequencies, modulation levels, coding rates, transmit optical powers}
\Hline{\hspace{-.6 cm}\hrulefill}
\Statex  \hspace{-4mm}\textbf{Stage 1:} Routing, Grooming, Ordering
\For{all connection requests $q$}
\While{$\mathcal{R}_q \geqslant \mathcal{C}$}
\State add a new full load transmit transponder at node $S(q)$;
\State add a new full load receive transponder at node $D(q)$;
\State use shortest path for the added transponder pair;
\State $\mathcal{R}_{q} \leftarrow \mathcal{R}_{q}-\mathcal{C}$;
\EndWhile
\EndFor
\State sort traffics $q=1, \cdots, \abs{\mathbf{Q}}$ decreasingly according to the product of shortest path length and rate;
\For{all traffics $q$}
\State $\text{MSPL} \leftarrow \infty$;
\State $\text{MATC} \leftarrow 0$;
\For{all groom scenarios on $q$th traffic shortest path}
\State $\text{MSPL}_t \leftarrow \text{MSPL of the current path decomposition}$;
\State $\text{MATC}_t \leftarrow \text{MATC of the current path decomposition}$;
\If{$\text{MATC}_t < \mathcal{R}_q$ }
\State \textbf{continue};
\EndIf
\If{$\text{MSPL}_t < \text{MSPL}$}
\State $\text{MSPL} \leftarrow \text{MSPL}_t$;
\State $\text{MATC} \leftarrow \text{MATC}_t$;
\EndIf
\EndFor
\State groom $q$th request to the scenario having $\text{MATC}$ and $\text{MSPL}$;
\State update sub-paths traffic volumes;
\State remove the groomed path;
\EndFor
\State assign transmit and receive transponder pairs to the remaining traffic requests;
\State route the remaining connection requests over their shortest path;
\State order all the assigned paths according to the product of shortest path length and rate;
\Statex  \hspace{-5mm}\textbf{Stage 2:} Transponder Parameter Assignment
\Do
\State use convex TPA to optimize transponder parameters;
\State $\mathcal{I}\leftarrow 0$;
\Do
\For{all transmit transponders $(v,t)$}
\For{all spectral efficiency values $c$ of Tab. \ref{tab:snr_spc}}
\If {if $c_{(v,t)}$ falls in $\mathcal{I}$-neighborhood of $c$}
\State round and fix $c_{(v,t)}$ to $c$;
\State \textbf{break};
\EndIf
\EndFor
\For{all coding rate values $r$ of Tab. \ref{tab:snr_spc}}
\If {if $r_{(v,t)}$ falls in $\mathcal{I}$-neighborhood of $r$}
\State round and fix $r_{(v,t)}$ to $r$;
\State \textbf{break};
\EndIf
\EndFor
\EndFor
\State $\mathcal{I} \leftarrow \mathcal{I}+\mathcal{J}$;
\doWhile{there is no valid value in $\mathbf{c}$ and $\mathbf{r}$}
\State remove fixed values of $\mathbf{c}$ and $\mathbf{r}$;
\doWhile{there is an inacceptable value in $\mathbf{c}$ and $\mathbf{r}$}
\end{algorithmic}
\end{algorithm}
\section{Routing, Grooming and Ordering Sub-Problem}\label{Sec_IV}
In this section, an ILP is formulated for routing and traffic grooming and then a heuristic approach is proposed to solve it.
\subsection{ILP Formulation}
Generally, traffic grooming aims at minimization of active transponders to save transponder power consumption while shortest path routing is usually used to minimize the length of routes and consequently the number of used optical amplifiers \cite{khodakarami2014flexible}. Routing and traffic grooming can be formulated as the following ILP:
\begin{align}
& \min_{\mathbf{x}, \mathbf{d}, \mathbf{f}, \mathbf{c}} \quad \sum\limits_{v \in \mathbf{V},l \in \mathbf{L},t \in \mathbf{T}}x_{(v,t,l)}\mathcal{L}_l+\mathcal{K}_1\sum\limits_{v,v' \in \mathbf{V},t,t' \in \mathbf{T}}d_{(v,t,v',t')}+\mathcal{K}_2\sum\limits_{v \in \mathbf{V},t \in \mathbf{T}}(\mathcal{C}-\sum\limits_{q \in \mathbf{Q}}f_{(v,t,q)})\label{eq:groom_g}\\
& \text{s.t.}\quad \sum\limits_{t \in \mathbf{T}}f_{(v,t,q)}+\delta(v-D(q))\mathcal{R}_q= \sum\limits_{t \in \mathbf{T}}c_{(v,t,q)}+\delta(v-S(q))\mathcal{R}_q,  \forall v \in \mathbf{V},  \forall q \in \mathbf{Q}\label{eq:groom_c1}\\
&\text{\quad} \sum\limits_{q \in \mathbf{Q}}f_{(v,t,q)} \leqslant \mathcal{C},  \forall v \in \mathbf{V}, \forall t \in \mathbf{T}\label{eq:groom_c2}\\
&\text{\quad} \sum\limits_{q \in \mathbf{Q}}c_{(v,t,q)} \leqslant \mathcal{C},  \forall v \in \mathbf{V}, \forall t \in \mathbf{T}\label{eq:groom_c3}\\
&\text{\quad} \abs{c_{(v',t',q)}-f_{(v,t,q)}} \leqslant \mathcal{K}(1-d_{(v,t,v',t')}),  \forall v,v' \in \mathbf{V}, \forall t,t' \in \mathbf{T}, \forall q \in \mathbf{Q}\label{eq:groom_c4}\\
&\text{\quad}\sum\limits_{q \in \mathbf{Q}}f_{(v,t,q)} \leqslant \mathcal{K}\sum\limits_{l \in \mathbf{L}: B(l)=v}x_{(v,t,l)},  \forall v \in \mathbf{V}, \forall t \in \mathbf{T}\label{eq:groom_c5}\\
&\text{\quad} \hspace{-0.5cm} \sum\limits_{l \in \mathbf{L}: E(l)=v'}x_{(v,t,l)} \leqslant \sum\limits_{l \in \mathbf{L}: B(l)=v'}x_{(v,t,l)} + \sum\limits_{t' \in \mathbf{T}}d_{(v,t,v',t')} \leqslant 1,  \forall v,v' \in \mathbf{V}, \forall t \in \mathbf{T}\label{eq:groom_c6}
\end{align}
where $\mathcal{K}$ is a large positive number. Each transmit transponder is assigned a route to its destination so, minimization of used link lengths automatically results in simultaneous shortest path routing and minimum number of active transponders. This is the main cost of the goal function. The second term of the goal function is a penalty term to prevent unnecessary add/drop of the traffic paths in grooming switches. Furthermore, we add another penalty term to incur a cost for transponders with unused remaining capacity and therefore, force the transponders to be fully utilized. Constraint \eqref{eq:groom_c1} is KCL law for added, dropped, input, output and groomed traffic of $q$th connection request at node $v$. Constraints \eqref{eq:groom_c2} and \eqref{eq:groom_c3} hold the information bit rate of the transmit and receive transponders below their capacity. Constraint \eqref{eq:groom_c4} states that all the traffics of a transmit transponder should be dropped at its assigned receive transponder. Constraint \eqref{eq:groom_c5} specifies that if a transponder has no allocated route, it should be inactive. Finally, constraint \eqref{eq:groom_c6} is the  optical switching constraint which states that if a traffic arrives to a node, it should be dropped to a receive transponder or switched to a neighbor node. This optimization problem has $2\abs{\mathbf{V}}\abs{\mathbf{T}}\abs{\mathbf{Q}}+\abs{\mathbf{V}}\abs{\mathbf{T}}\abs{\mathbf{L}}+\abs{\mathbf{V}}^2\abs{\mathbf{T}}^2$ integer variables and its number of constraints has the order of $\abs{\mathbf{V}}^2\abs{\mathbf{T}}^2\abs{\mathbf{Q}}$. Although the problem is ILP, its number of variables and constraints is very large for practical values of $\abs{\mathbf{V}}$, $\abs{\mathbf{T}}$ and $\abs{\mathbf{Q}}$. Therefore, a heuristic approach is required to provide a fast-achievable near-optimum solution for routing and traffic grooming problem.

\subsection{Heuristic Solution}
\begin{figure}[t!] 
\center{\includegraphics[scale=0.55]{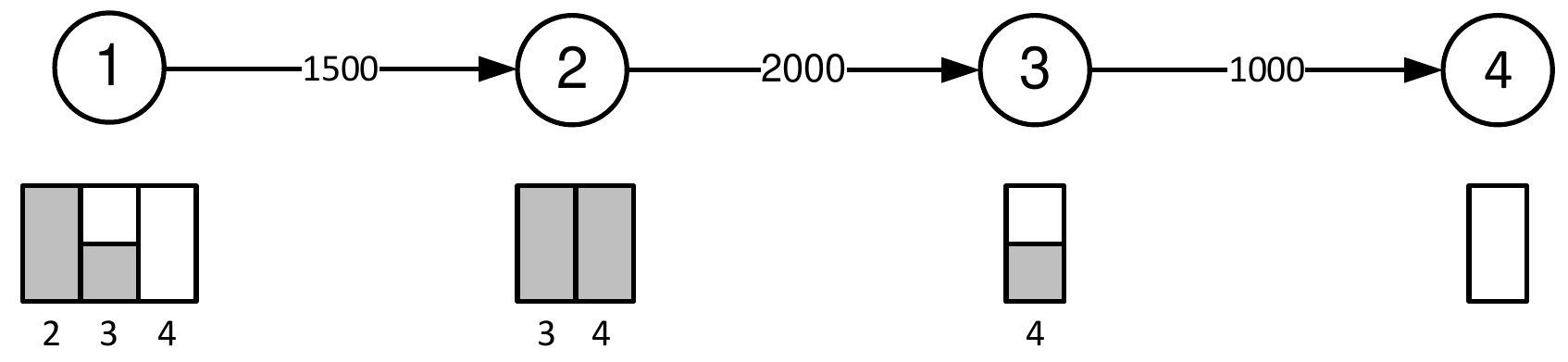}}
\center{\caption{\label{fig:groom} An example for traffic grooming.}}
\end{figure}

Long and high volume traffic requests are more susceptible to amplifier noise and nonlinear interference effects and need low level modulations, low coding rates, high transmit optical powers and therefore consume a lot of power. If such traffics are groomed to short traffic segments, each segment can be routed using high modulation levels and coding rates and consequently consumes lower spectrum and power. Merging long and high volume traffics to existing short  traffic requests also reduces the number of required transponders and assigned links by utilizing the unused remaining capacity of the transponders which is our goal in ILP of \eqref{eq:groom_g}-\eqref{eq:groom_c6}. It is conventionally stated that if the path length is reduced by a factor of $0.5$, modulation level can be doubled \cite{chatterjee2015routing}. This in turn reduces the assigned spectrum width by a factor of $0.5$ and results in lower transponder power consumption according to \eqref{eq:trx_pow}. Consider the sample scenario of Fig. \ref{fig:groom} in which transmit transponders are shown by rectangles below the nodes. For each transponder, the used capacity is highlighted by dark color and the number below each transponder shows its destination node. Assume we should route an input traffic with rate $\frac{\mathcal{C}}{2}$ from source $1$ to destination $4$ over its shortest path which is $1\rightarrow 2\rightarrow 3\rightarrow 4$. There are four grooming scenarios over this path which are summerized in Tab. \ref{tab:groom}. Minimum Available Transponder Capacity (MATC) is defined as the remaining capacity of the transmit transponder that has the maximum load over sub-paths of the grooming scenario. For example in scenario $3$, we should use transponder $3$ of node $1$ and transponder $4$ of node $3$ which both of them have remaining capacity of $\frac{\mathcal{C}}{2}$, therefore $\text{MATC} = \frac{\mathcal{C}}{2}$. Maximum Sub Path Length (MSPL) is the maximum length of the sub-paths along the grooming scenario. For example in scenario $3$, the sub-path $1\rightarrow 3$ has the maximum length which is $3500$. As the value of MSPL decreases, there are more chances to partition the path to shortest sub-paths and save power. As reported in Tab. \ref{tab:groom}, $1\rightarrow 3\rightarrow 4$ has the minimum MSPL and its MATC is enough to accommodate the required load and therefore, it is our chosen grooming scenario. Note that at worst case the traffic is routed over its shortest path without grooming to sub-paths.

During RGO stage of Alg. \ref{alg:rsa}, traffic volumes are partitioned such that no traffic volume is greater than transponder capacity $\mathcal{C}$. Then, we choose partitioned traffic volumes with full load of $\mathcal{C}$ and route them over their shortest paths. The remaining traffic volumes are descendingly sorted according to the product of traffic volume and shortest path length. Next, we groom the sorted traffics using the procedure explained above, as written in Alg. \ref{alg:rsa}. Finally, we use the metric of rate-length product to order traffics over links as explained in \cite{hadi2017resource}.
\begin{table}[t!]
\small
\centering
\caption{Grooming scenarios for example of Fig. \ref{fig:groom}}.\label{tab:groom}
\begin{tabular}{cccc}
\hline
\hline
Scenario Number & Grooming Scenario & MATC & MSPL \\
\hline
$1$ & $1\rightarrow 4$ & $\mathcal{C}$ & $4500$\\
$2$ & $1\rightarrow 2\rightarrow 4$ & $0$ & $3000$\\
$3$ & $1\rightarrow 3\rightarrow 4$ &  $\frac{\mathcal{C}}{2}$ & $3500$\\
$4$ & $1\rightarrow 2\rightarrow 3\rightarrow 4$ &  $0$ & $2000$\\
\hline
\hline
\end{tabular}
\end{table}

\section{Transponder Parameter Assignment Sub-Problem}\label{Sec_V}
In this section, we provide an MINLP formulation for TPA and then show how it can be converted to a convex optimization problem.
\subsection{Nonlinear Formulation}\label{Sec_V_A}
A MINLP formulation for TPA is as follows \cite{yan2015resource}:
\begin{align}
&\min_{\mathbf{c}, \mathbf{b}, \mathbf{r}, \mathbf{p}, \bm{\omega}} \quad \sum\limits_{v \in \mathbf{V}, t \in \mathbf{T}}X_{(v,t)}\label{eq:nonlinear_g}\\
&\text{s.t.} \quad \Psi_{(v,t)} \geqslant \Theta_{(v,t)}, \quad \forall v \in \mathbf{V}, \forall t \in \mathbf{T} \label{eq:nonlinear_c1}\\
& \text{\quad} \omega_{\Upsilon_{l, j}}+\frac{\Delta_{\Upsilon_{l, j}}}{2} + \mathcal{G}\leqslant \omega_{\Upsilon_{l, j+1}}-\frac{\Delta_{\Upsilon_{l, j+1}}}{2}, \quad \forall l \in \mathbf{L}, j = 1, \cdots, \abs{\mathbf{H}_l}-1 \label{eq:nonlinear_c2}\\
&\text{\quad} \frac{\Delta_{(v,t)}}{2} \leqslant \omega_{(v,t)} \leqslant \mathcal{B}- \frac{\Delta_{(v,t)}}{2}, \quad \forall v \in \mathbf{V}, \forall t \in \mathbf{T} \label{eq:nonlinear_c3}\\
&\text{\quad} \sum\limits_{q}f_{(v,t,q)} \leqslant 2 r_{(v,t)} c_{(v,t)}\Delta_{(v,t)}, \quad \forall v \in \mathbf{V}, \forall t \in \mathbf{T} \label{eq:nonlinear_c4}
\end{align}
where $\mathbf{c}$, $\mathbf{b}$, $\mathbf{r}$, $\mathbf{p}$ and $\bm{\omega}$ are variable vectors of transponder configuration parameters \ie modulation level, number of sub-carriers, coding rate, transmit optical power and central frequency. The goal is to minimize the total transponder power consumption where $X_{(v,t)}$ is obtained using \eqref{eq:trx_pow}. Constraint \eqref{eq:nonlinear_c1} is the QoS constraint that forces $\Psi_{(v,t)}$, the OSNR of transponder $(v,t)$, to be greater than its required minimum threshold $\Theta_{(v,t)}$. $\Psi_{(v,t)}$ is a nonlinear function of $\mathbf{b}$, $\mathbf{p}$ and $\bm{\omega}$ \cite{hadi2017resource, yan2017joint} while the value of $\Theta_{(v,t)}$ are picked up from Tab. \ref{tab:snr_spc}. Constraint \eqref{eq:nonlinear_c2} is nonoverlapping-guard constraint that prevents two transmit transponders share a same frequency spectrum. $\Upsilon_{l, j}$ is a function that shows which transponder occupies $j$-th spectrum bandwidth on link $l$ and its values are determined during ordering part of RGO sub-problem. Constraint \eqref{eq:nonlinear_c3} holds all assigned central frequencies within the acceptable range of the fiber spectrum and finaly the last constraint guarantees that the transponder can convey the input traffic rate $\sum_{q}f_{(v,t,q)}$.  In fact, this MINLP is an extension of the formulation of \cite{khodakarami2014flexible} in which transmit optical powers also contribute to the optimization and QoS constraint is specified by the general and nonlinear expressions of \cite{yan2015resource}.
\subsection{Convexification Technique}\label{Sec_V_B}
\begin{figure}[t!] 
\center{\includegraphics[scale=0.6]{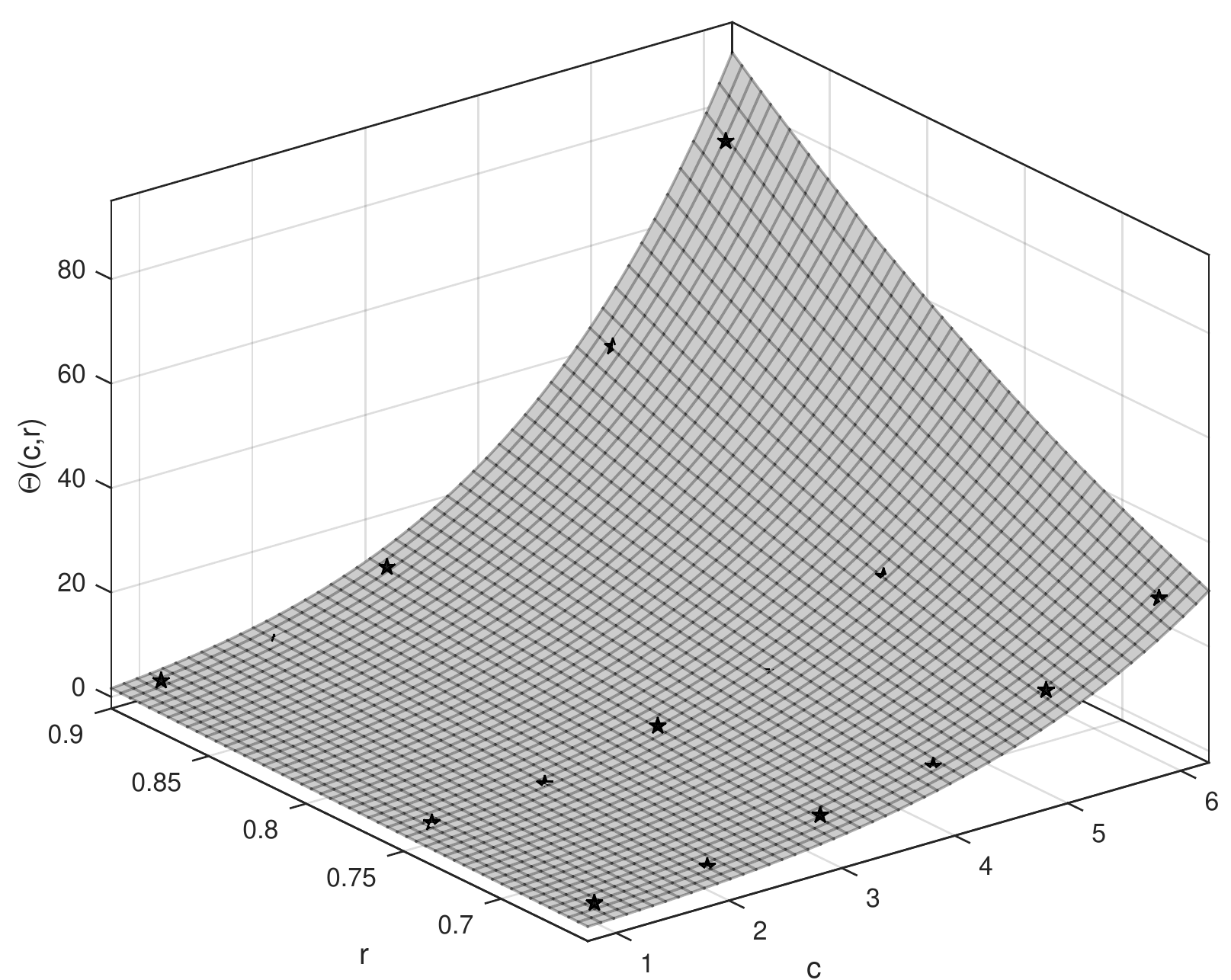}}
\center{\caption{\label{fig:snr} 2D curve fitting for $(c,r, \Theta)$ data samples given in Tab. \ref{tab:snr_spc}.}}
\end{figure}
We first provide a generalized posynomial expression \cite{boyd2007tutorial} for the optimization and then define a variable change to convexify the problem. A posynomial expression for OSNR has been proposed in \cite{hadi2017resource}:
\begin{align}\label{eq:xci_app}
 \Psi_{(v,t)}=  \frac{p_{(v,t)}}{\zeta \mathcal{N}_{(v,t)}\Delta_{(v,t)}+\kappa_1\varsigma p_{(v,t)}\sum\limits_{\substack{v' \in \mathbf{V}, t' \in \mathbf{T}\\ (v',t') \neq (v,t)}}\frac{p_{(v',t')}^{2}\mathcal{N}_{(v,t,v',t')}}{\Delta_{(v',t')}d_{(v,t,v',t')}}+\varsigma\iota\mathcal{N}_{(v,t)}p_{(v,t)}^3}, \forall v \in \mathbf{V}, \forall t \in \mathbf{T}
\end{align}
where $\kappa_1=0.4343$, $\zeta=(e^{\alpha \mathcal{L}_{spn}}-1)h\nu n_{sp}$, $\varsigma=\frac{3\gamma^2}{2\alpha\pi\abs{\beta_2}}$ and $\iota=\frac{\pi^2\abs{\beta_2}}{2\alpha}$. $n_{sp}$ is the spontaneous emission factor, $\nu$ is the light frequency, $h$ is Planck’s constant, $\alpha$ is attenuation coefficient, $\beta_2$ is dispersion factor and $\gamma$ is nonlinear constant. Furthermore,  $d_{(v,t,v',t')}$ is the distance between carrier frequencies $\omega_{(v,t)}$ and $\omega_{(v',t')}$ and equals to $d_{(v,t,v',t')} = \abs{\omega_{(v,t)}-\omega_{(v',t')}}$. Similar to \cite{hadi2017resource}, we use a posynomial curve fitting for date samples of Tab. \ref{tab:snr_spc}:
\begin{align} \label{eq:snr_osnr_app}
\Theta(c, r) \approx r^{\kappa_2}(1+\kappa_3 c)^{\kappa_4}, 1 \leqslant c \leqslant 6,   0.6 \leqslant r \leqslant 1
\end{align}
where $\kappa_2=3.37$, $\kappa_3=0.21$, $\kappa_4=5.73$. Fig. \ref{fig:snr} shows data samples and the posynomial expression for data curve fitting.

Following the same approach as \cite{hadi2017resource}, we arrive at this new representation of the optimization problem:
\begin{align}
&\min_{\mathbf{c}, \mathbf{b}, \mathbf{r}, \mathbf{p}, \bm{\omega}, \mathbf{t}, \mathbf{d}} \quad \sum\limits_{v \in \mathbf{V}, t \in \mathbf{T}}X_{(v,t)}+\mathcal{K}_3\sum\limits_{\substack{v,v' \in \mathbf{V}, t,t' \in \mathbf{T}\\(v,t) \neq (v',t')\\ \mathcal{N}_{(v,t,v',t')} \neq 0}}d_{(v,t,v',t')}^{-1}\label{eq:gp_1_g}\\
\nonumber &\text{s.t.} \quad  r_{(v,t)}^{\kappa_2} t_{(v,t)}^{\kappa_4}\Big[\zeta\mathcal{F}\mathcal{N}_{(v,t)}p_{(v,t)}^{-1}2^{b_{(v,t)}}+\varsigma\iota \mathcal{N}_{(v,t)}p_{(v,t)}^2+ \kappa_1\varsigma\mathcal{F}^{-1}\sum\limits_{\substack{v' \in \mathbf{V}, t' \in \mathbf{T}\\ (v',t') \neq (v,t)}}p_{(v',t')}^{2}\mathcal{N}_{(v,t,v',t')}2^{-b_{(v',t')}} d_{(v,t,v',t')}^{-1}\Big]\\
& \text{\quad}\leqslant 1 , \quad \forall v \in \mathbf{V}, \forall t \in \mathbf{T} \label{eq:gp_1_c1} \\
& \text{\quad} \omega_{\Upsilon_{l, j}}\omega_{\Upsilon_{l, j+1}}^{-1}+0.5\mathcal{F}2^{b_{\Upsilon_{l, j}}} \omega_{\Upsilon_{l, j+1}}^{-1}+\mathcal{G}\omega_{\Upsilon_{l, j+1}}^{-1}+ 0.5\mathcal{F}2^{b_{\Upsilon_{l, j+1}}} \omega_{\Upsilon_{l, j+1}}^{-1}\leqslant 1, \quad \forall l \in \mathbf{L}, j = 1, \cdots, \abs{\mathbf{H}_l}-1 \label{eq:gp_1_c2}\\
&\text{\quad} 0.5\mathcal{F}2^{b_{(v,t)}}\mathcal{B}^{-1} + \omega_{(v,t)}\mathcal{B}^{-1} \leqslant 1, \quad \forall v \in \mathbf{V}, \forall t \in \mathbf{T} \label{eq:gp_1_c3} \\
& \text{\quad} 0.5\mathcal{F}2^{b_{(v,t)}}\omega_{(v,t)}^{-1} \leqslant 1, \quad \forall v \in \mathbf{V}, \forall t \in \mathbf{T}\label{eq:gp_1_c4}  \\
& \text{\quad} 0.5\mathcal{F}^{-1} r_{(v,t)}^{-1}  c_{(v,t)}^{-1} 2^{-b_{(v,t)}}\sum\limits_{q}f_{(v,t,q)}\leqslant 1, \quad \forall v \in \mathbf{V}, \forall t \in \mathbf{T} \label{eq:gp_1_c5}  \\
& \text{\quad} t_{(v,t)}^{-1}+\kappa_3c_{(v,t)}t_{(v,t)}^{-1} \leqslant 1, \quad \forall v \in \mathbf{V}, \forall t \in \mathbf{T}  \label{eq:gp_1_c6}  \\
&\text{\quad}  d_{(\Upsilon_{l, i},\Upsilon_{l, j})} \omega_{\Upsilon_{l, i}}^{-1}+ \omega_{\Upsilon_{l, j}}\omega_{\Upsilon_{l, i}}^{-1}\leqslant  1, \forall l \in \mathbf{L}, j = 1, \cdots, \abs{\mathbf{H}_l}-1, i = j+1, \cdots, \abs{\mathbf{H}_l}\label{eq:gp_1_c7}\\
&\text{\quad} d_{(\Upsilon_{l, i},\Upsilon_{l, j})}\omega_{\Upsilon_{l, j}}^{-1}+\omega_{\Upsilon_{l, i}}\omega_{\Upsilon_{l, j}}^{-1} \leqslant  1 ,\quad \forall l \in \mathbf{L}, j = 2, \cdots, \abs{\mathbf{H}_l}, i = 1, \cdots, j-1 \label{eq:gp_1_c8} 
\end{align}
Ignoring constraints \eqref{eq:gp_1_c6}, \eqref{eq:gp_1_c7}, \eqref{eq:gp_1_c8} and the penalty term of the goal function \eqref{eq:gp_1_g}, the above formulation is equivalent GP of the previous MINLP in which expressions \eqref{eq:xci_app} and \eqref{eq:snr_osnr_app} have been used for QoS constraint \eqref{eq:gp_1_c1}. Constraints \eqref{eq:gp_1_c6} and \eqref{eq:gp_1_c7} and the penalty term are added to guarantee the implicit equality of $d_{(v,t,v',t')} = \abs{\omega_{(v,t)}-\omega_{(v',t')}}$ \cite{hadi2017resource}. Constraint \eqref{eq:gp_1_c6} is also needed to convert the generalized posynomial QoS constraint to a valid GP one, as explained in \cite{boyd2007tutorial}. Now, consider the following variable change:
\begin{align}\label{eq:vc}
\nonumber & x=e^{X}:x \in \mathbb{R}_{>0} \longrightarrow X \in \mathbb{R}, \quad \forall x \in \mathbf{c} \cup \mathbf{r} \cup \mathbf{p}\cup  \bm{\omega} \cup  \mathbf{t} \cup  \mathbf{d}\\
& x=X: x\in\mathbb{R}_{>0}\longrightarrow X \in \mathbb{R}_{>0}, \quad \forall x \in \mathbf{b}
\end{align}
If one applies this variable change, the new representation of the problem is convex. As an example for the goal function we have:
\begin{align}\label{eq:cv_g}
 \sum\limits_{v \in \mathbf{V}, t \in \mathbf{T}} [\mathcal{P}_{txb}+\mathcal{P}_{rxb}+(\mathcal{P}_{dec}+\mathcal{P}_{enc})e^{-r_{(v,t)}}+b_{(v,t)}2^{b_{(v,t)}}\mathcal{P}_{fft}+2^{b_{(v,t)}} \mathcal{P}_{dsp}]+\mathcal{K}\sum\limits_{\substack{v,v' \in \mathbf{V}, t,t' \in \mathbf{T}\\(v,t) \neq (v',t')\\ \mathcal{N}_{(v,t,v',t')} \neq 0}}e^{-d_{(v,t,v',t')}}
\end{align}
Clearly, $e^{-d_{(v,t,v',t')}}$ and $e^{-r_{(v,t)}}$ are convex over $\mathbb{R}$ while $2^{b_{(v,t)}} $ and $b_{(v,t)}2^{b_{(v,t)}}$ are convex over $\mathbb{R}_{>0}$.  Consequently, function \eqref{eq:cv_g} which is a nonnegative weighted sum of convex functions is also convex. The same statement can be applied to show the convexity of the constraints under variable change of \eqref{eq:vc} (for some constraints, we need to apply an extra $\log$ to both side of the inequality to have a convex constraint). 

The original MINLP has $5\abs{\mathbf{V}}\abs{\mathbf{T}}$ variables and at most $4\abs{\mathbf{V}}\abs{\mathbf{T}} + \abs{\mathbf{V}}\abs{\mathbf{T}}\abs{\mathbf{L}}$ constraints while its convex representation requires  $6\abs{\mathbf{V}}\abs{\mathbf{T}}+\abs{\mathbf{V}}^2\abs{\mathbf{T}}^2$ variables and at most $5\abs{\mathbf{V}}\abs{\mathbf{T}} + 3\abs{\mathbf{V}}\abs{\mathbf{T}}\abs{\mathbf{L}}$. Also the new convex representation of the problem has more variables and constraints than its MINLP counterpart but it can be quickly solved using fast convex optimization algorithms. In the second stage of Alg. \ref{alg:rsa}, a relaxed continuous version of the proposed mixed-integer convex formulation is iteratively optimized in a local loop \cite{boyd2007tutorial}. At each epoch, the continuous convex optimization is solved and obtained values for relaxed integer variables (here, $c_{(v,t)}$'s and $r_{(v,t)}$'s) are rounded by precision $\mathcal{I}$. If none of the rounded variables is valid based on Tab. \ref{tab:snr_spc}, the precision is increased by $\mathcal{J}$ and the rounding is again applied. This process continues until at least one valid element is found for relaxed integer variables. Then, we fix the acceptable rounded values and solve the relaxed continuous convex problem again. The local loop continues until all the integer variables have valid and acceptable values. Note that the number of iterations in the heuristic algorithm is at most equal to the number of integer variables (here, at most $2\abs{\mathbf{V}}\abs{\mathbf{T}}$). The number of iterations is usually lower than its maximum. Furthermore, a simpler problem should be solved as the number of iteration increases because of fixing some of the integer variables during each loop.

\section{Numerical Results}\label{Sec_VI}
\begin{figure}[t!] 
\center{\includegraphics[scale=0.5]{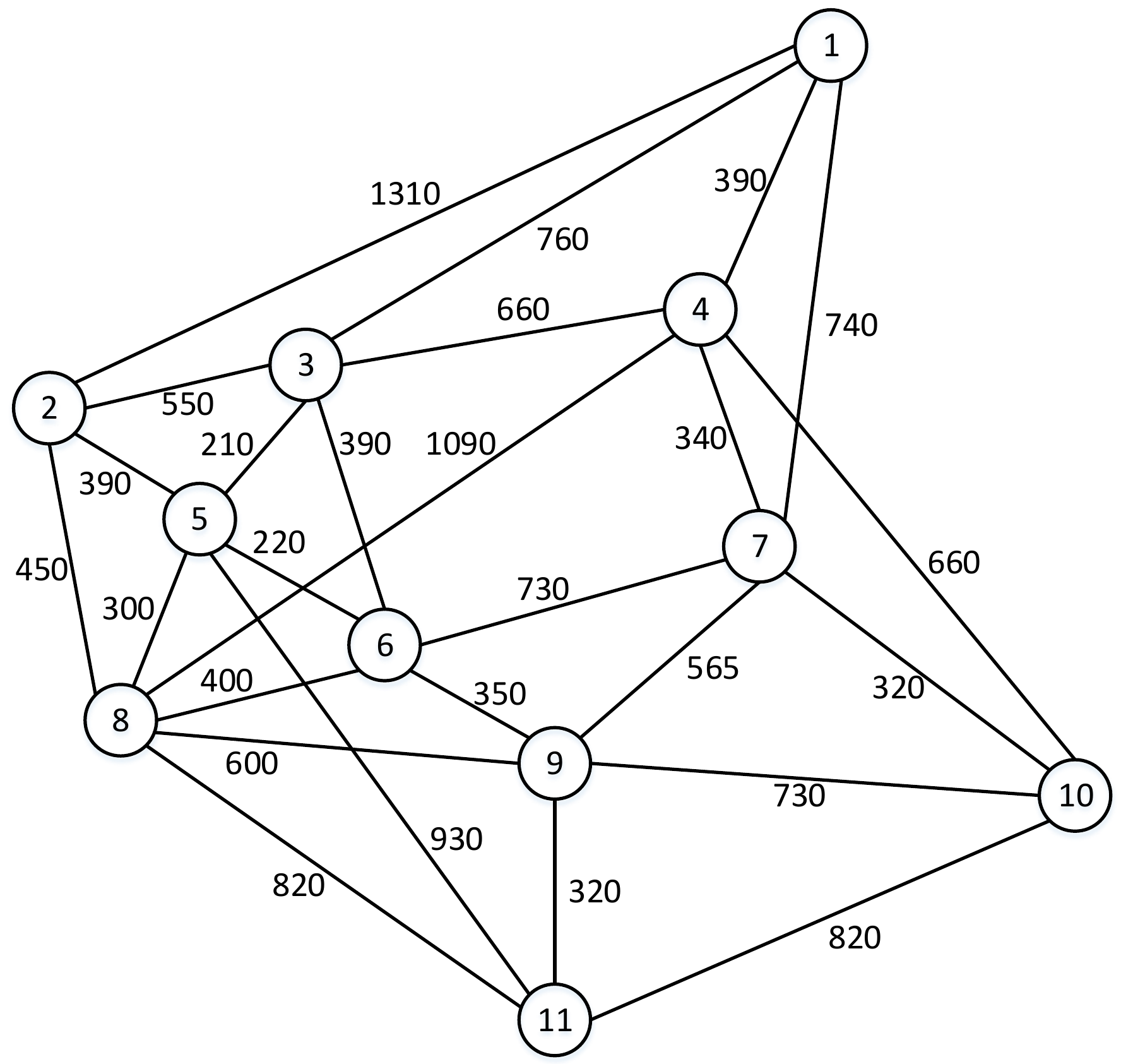}}
\center{\caption{\label{fig:euronet} European Cost239 optical network with $11$ optical nodes and $26$ bi-directional links. Digits on each link shows its length in km.}}
\end{figure}

In this section, we use simulation results to demonstrate the performance of the proposed formulations and algorithms. The European Cost239 optical network is considered with the topology of Fig. \ref{fig:euronet} and the normalized traffic matrix given in Tab. \ref{tab:traffic_matrix}. One can scale the normalized traffic values of Tab. \ref{tab:traffic_matrix} to get $2017$ west Europe aggregate traffic value of $16.8$ EByte/month \cite{muhammad2014introducing, khodakarami2014flexible}. The optical link parameters and other simulation constants are reported in Tab. \ref{tab:sim_param}. We use MATLAB software for programming and simulation \cite{matlabmathworks}. YALMIP \cite{lofberg2005yalmip} and CVX \cite{grant2008cvx} optimization applications  are also used for modeling and solving the proposed optimization problems. Simulations run over a computer equipped with Intel Core i7-472HQ CPU and 8 GB of RAM.
\begin{table*}[t!]
\tiny
\centering
\caption{Simulation constant parameters.}\label{tab:sim_param}
\begin{tabular}{ccccccccccccccccccc}
\hline
\hline
$\abs{\substack{\beta_2}}$ & $\alpha$ & $L$ & $\nu$ & $n_{sp}$ & $\gamma$ & $\mathcal{G}$ & $\mathcal{B}$  & $\mathcal{J}$  & $\mathcal{P}_{txb}$ & $\mathcal{P}_{rxb}$ & $\mathcal{P}_{enc}$ & $\mathcal{P}_{dec}$ & $\mathcal{P}_{fft}$ & $\mathcal{P}_{dsp}$ &  $\mathcal{E}_{grm}$ & $\mathcal{P}_{amp}$\\
$\text{fs}^2/\text{m}$ & dB/km & km & THz &  & 1/W/km & GHz & THz  & & W & W & W & W & mW & mW & pJ/bit & W\\
\hline
$20393$ & $0.22$ & $80$ & $193.55$ & $1.58$ & $1.3$ & $20$ & $2$ & $0.1$ & $16$ & $20$ & $0.2$ & $3$ & $4$ & $10$ & $400$ & $12$\\
\hline
\hline
\end{tabular}
\end{table*}
\begin{table}[t!]
\small
\centering
\caption{European Cost239 optical network normalized traffic matrix.}\label{tab:traffic_matrix}
\begin{tabular}{cccccccccccc}
\hline
\hline
\textbf{Node} & $\mathbf{01}$ & $\mathbf{02}$ & $\mathbf{03}$ & $\mathbf{04}$ & $\mathbf{05}$ & $\mathbf{06}$ & $\mathbf{07}$ & $\mathbf{08}$ & $\mathbf{09}$ & $\mathbf{10}$ & $\mathbf{11}$\\
\hline
$\mathbf{01}$ & $00$ & $01$ & $01$ & $03$ & $01$ & $01$ & $01$ & $35$ & $01$ & $01$ & $01$\\
$\mathbf{02}$ & $01$ & $00$ & $05$ & $14$ & $40$ & $01$ & $01$ & $10$ & $03$ & $02$ & $03$\\
$\mathbf{03}$ & $01$ & $05$ & $00$ & $16$ & $24$ & $01$ & $01$ & $05$ & $03$ & $01$ & $02$\\
$\mathbf{04}$ & $03$ & $14$ & $16$ & $00$ & $06$ & $02$ & $02$ & $21$ & $81$ & $09$ & $09$\\
$\mathbf{05}$ & $01$ & $40$ & $24$ & $06$ & $00$ & $01$ & $11$ & $06$ & $11$ & $01$ & $02$\\
$\mathbf{06}$ & $01$ & $01$ & $01$ & $02$ & $01$ & $00$ & $01$ & $01$ & $01$ & $01$ & $01$\\
$\mathbf{07}$ & $01$ & $01$ & $01$ & $02$ & $11$ & $01$ & $00$ & $01$ & $01$ & $01$ & $01$\\
$\mathbf{08}$ & $35$ & $10$ & $05$ & $21$ & $06$ & $01$ & $01$ & $00$ & $06$ & $02$ & $05$\\
$\mathbf{09}$ & $01$ & $03$ & $03$ & $81$ & $11$ & $01$ & $01$ & $06$ & $00$ & $51$ & $06$\\
$\mathbf{10}$ & $01$ & $02$ & $01$ & $09$ & $01$ & $01$ & $01$ & $02$ & $51$ & $00$ & $81$\\
$\mathbf{11}$ & $01$ & $03$ & $02$ & $09$ & $02$ & $01$ & $01$ & $05$ & $06$ & $81$ & $00$\\
\hline
\hline
\end{tabular}
\end{table}

\begin{figure}[t!] 
\center{\includegraphics[scale=0.5]{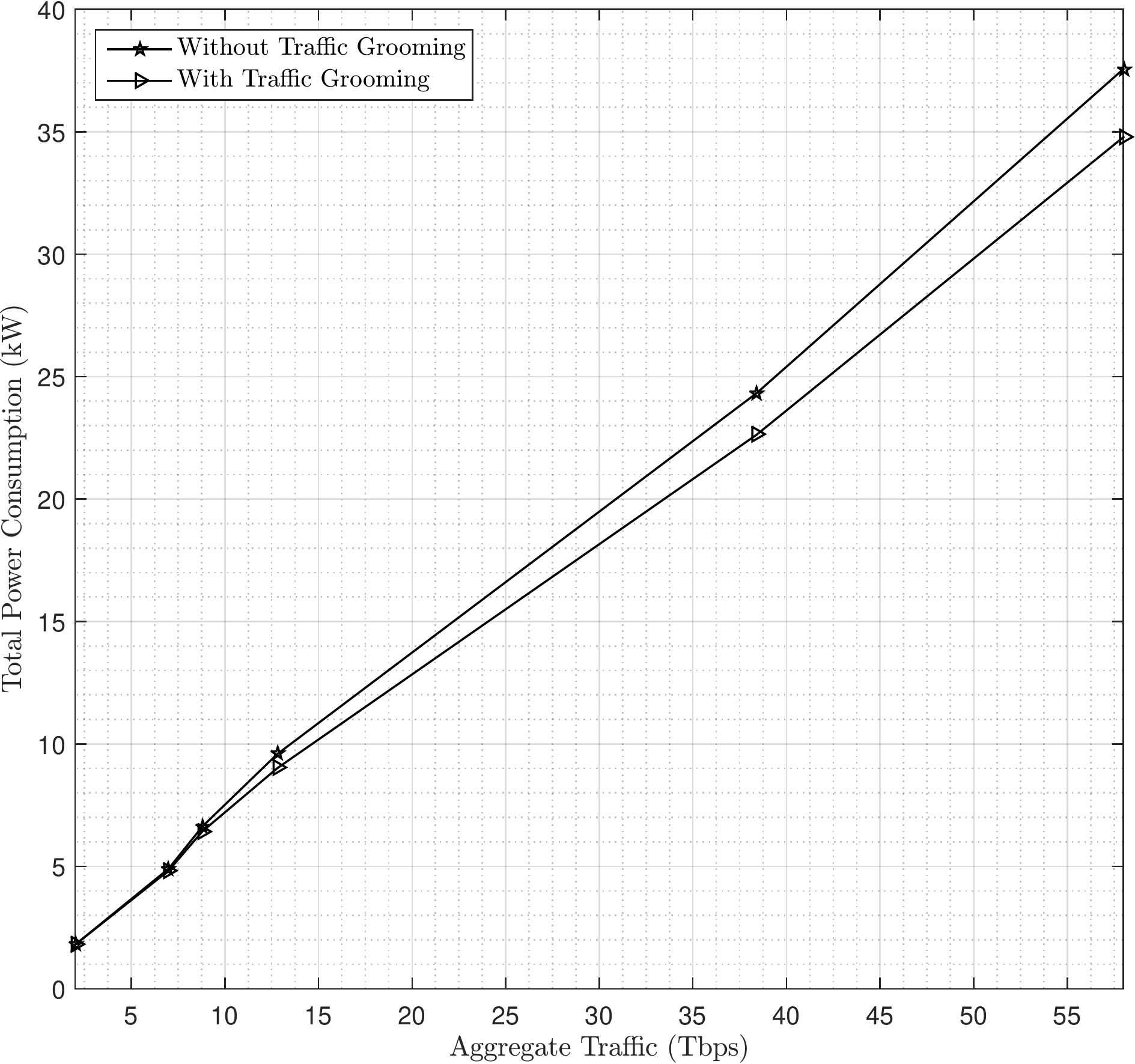}}
\center{\caption{\label{fig:groomPower} Total power consumption versus aggregate traffic with and without applying Alg. \ref{alg:rsa} for traffic grooming.}}
\end{figure}

Fig. \ref{fig:groomPower} compares total power consumption versus aggregate network traffic with and without traffic grooming procedure of Alg. \ref{alg:rsa}. Clearly, traffic grooming  reduces the consumed power and the amount of saved power is improved for higher aggregate traffic values where more traffic grooming scenarios are available. As an example, the power consumption is reduced by a factor of $9\%$ when aggregate traffic is about $60$ Tbps. 

\begin{figure}[t!] 
\center{\includegraphics[scale=0.5]{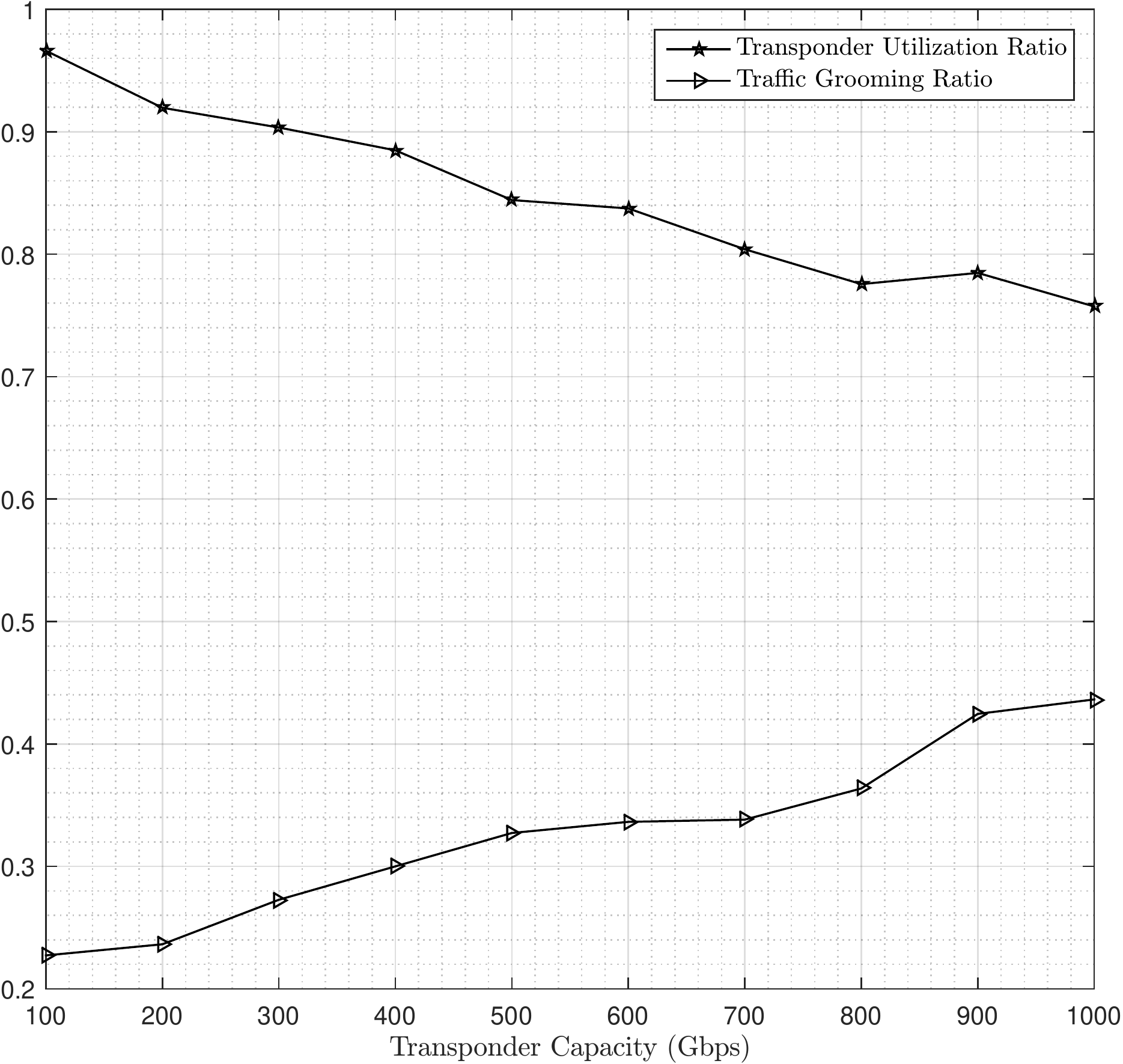}}
\center{\caption{\label{fig:trxCap} Transponder utilization and traffic grooming ratios versus transponder capacity $\mathcal{C}$.}}
\end{figure}

Alg. \ref{alg:rsa} partitions the input traffic streams according to the capacity of transponders $\mathcal{C}$  and uses the proposed traffic grooming procedure to fill transponders with remaining unused capacity. We define Transponder Utilization Ratio (TUR) as the ratio of the aggregate traffic to the total capacity offered by all transponders. $\text{TUR}=1$ means that all the transponders work at full capacity. Traffic Grooming Ratio (TGR) is defined as the ratio of the number of unloaded transponders by traffic grooming to the number of unfilled transponders when no traffic grooming is applied. TGR shows how traffic grooming reduces the number of required transponders. Fig. \ref{fig:trxCap} shows TUR and TGR versus transponder capacity $\mathcal{C}$. As $\mathcal{C}$ increases, there are more transponders with unused remaining capacity that can be used for traffic grooming and therefore, TGR is improved. On the other hand, we need more traffic volumes to use full capacity of the transponders when  $\mathcal{C}$ is increased and therefore, TUR goes down. Higher values of TUR reduces the required CAPEX for network deployment while higher values of TGR reduces the power consumption and consequently reduces the OPEX. Fig. \ref{fig:trxCap} shows that there is a tradeoff between TUR and TGR or equivalently CAPEX and OPEX for higher values of $\mathcal{C}$.  

\begin{figure}[t!] 
\center{\includegraphics[scale=0.5]{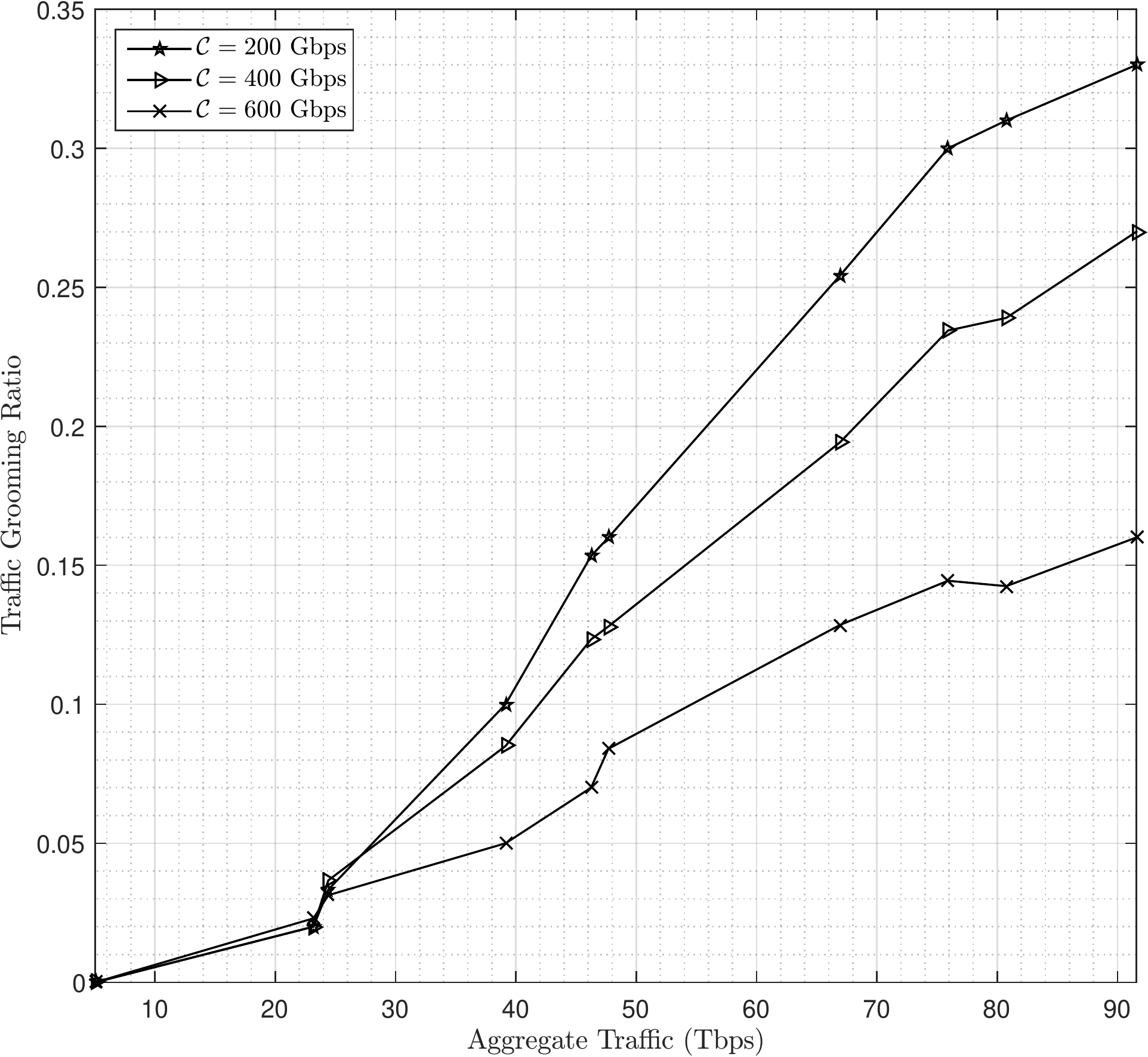}}
\center{\caption{\label{fig:groomTrxCap} Traffic grooming ratio versus aggregate network traffic for different values of transponder capacity. }}
\end{figure}

Fig. \ref{fig:groomTrxCap} shows TGR versus aggregate network traffic for different values of transponder capacity. As the aggregate traffic increases, more traffics are groomed and we save more power due to unloaded transponders. The improvement is more for transponders with higher capacity. As an example, when the aggregate traffic is $67$ Tbps and $\mathcal{C}$ increases from $200$ Gbps to $400$ Gbps, TGR is improved by a factor of $1.51$.

\begin{figure}[t!] 
\center{\includegraphics[scale=0.5]{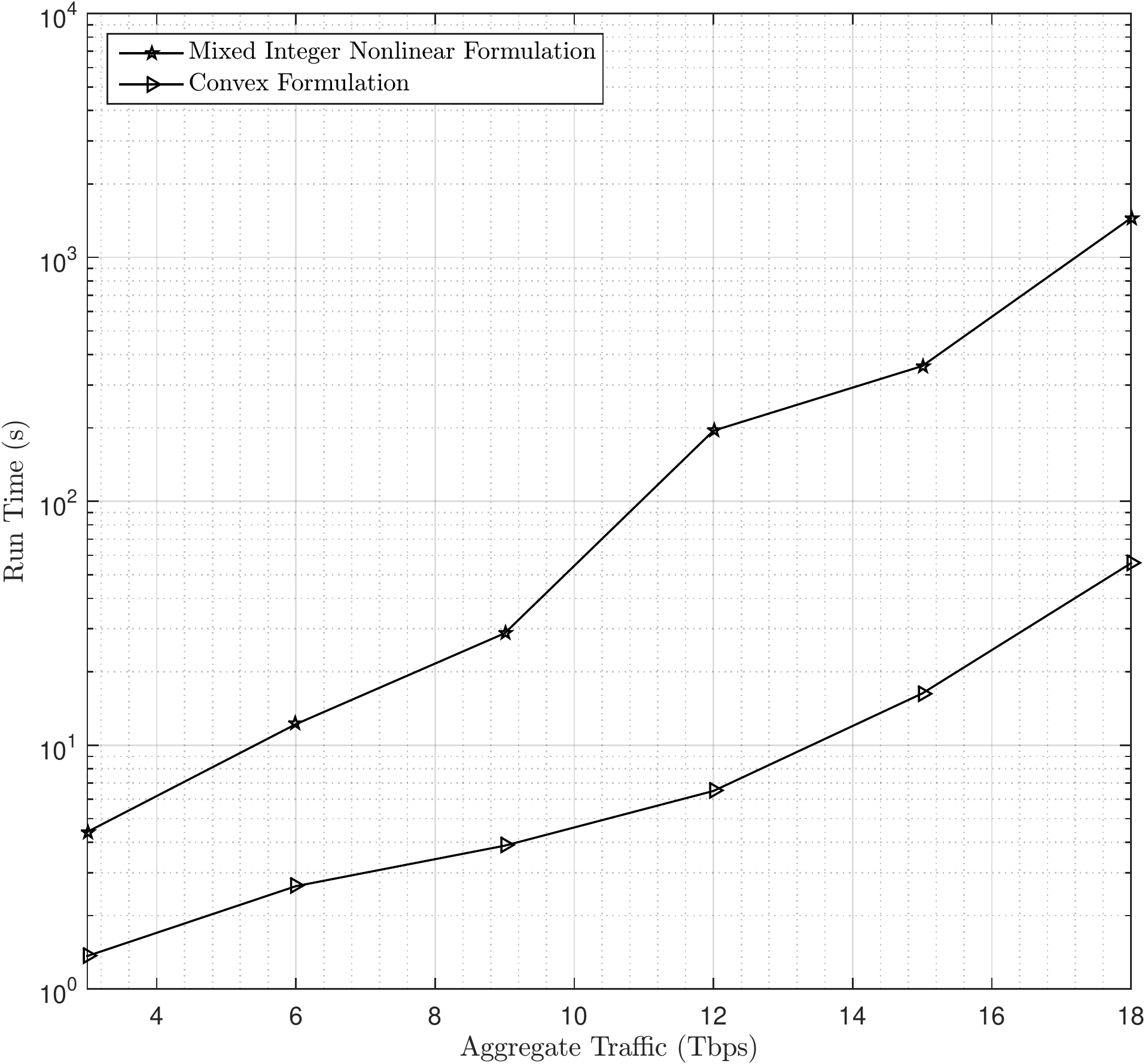}}
\center{\caption{\label{fig:time}Run time in terms of aggregate traffic for MINLP and convex formulations of transponder parameter assignment problem.}}
\end{figure}

Fig. \ref{fig:time} shows how our proposed TPA convex formulation runs faster than its MINLP counterpart. We use MINLP formulations of \eqref{eq:nonlinear_g}-\eqref{eq:nonlinear_c4} as a test bench in which QoS constraint is replaced by the nonlinear and general expression of \cite{yan2015resource}. Results show that the convex formulation is much faster than the MINLP formulation. As an example, it can be about $20$ times faster at aggregate traffic of $18$ Tbps. Simulation results also show that for higher values of aggregate traffic, the nonlinear optimizer package may fail to provide a feasible solution while the convex optimizer software still yields a feasible solution in a reasonable run time.

\begin{figure}[t!] 
\center{\includegraphics[scale=0.5]{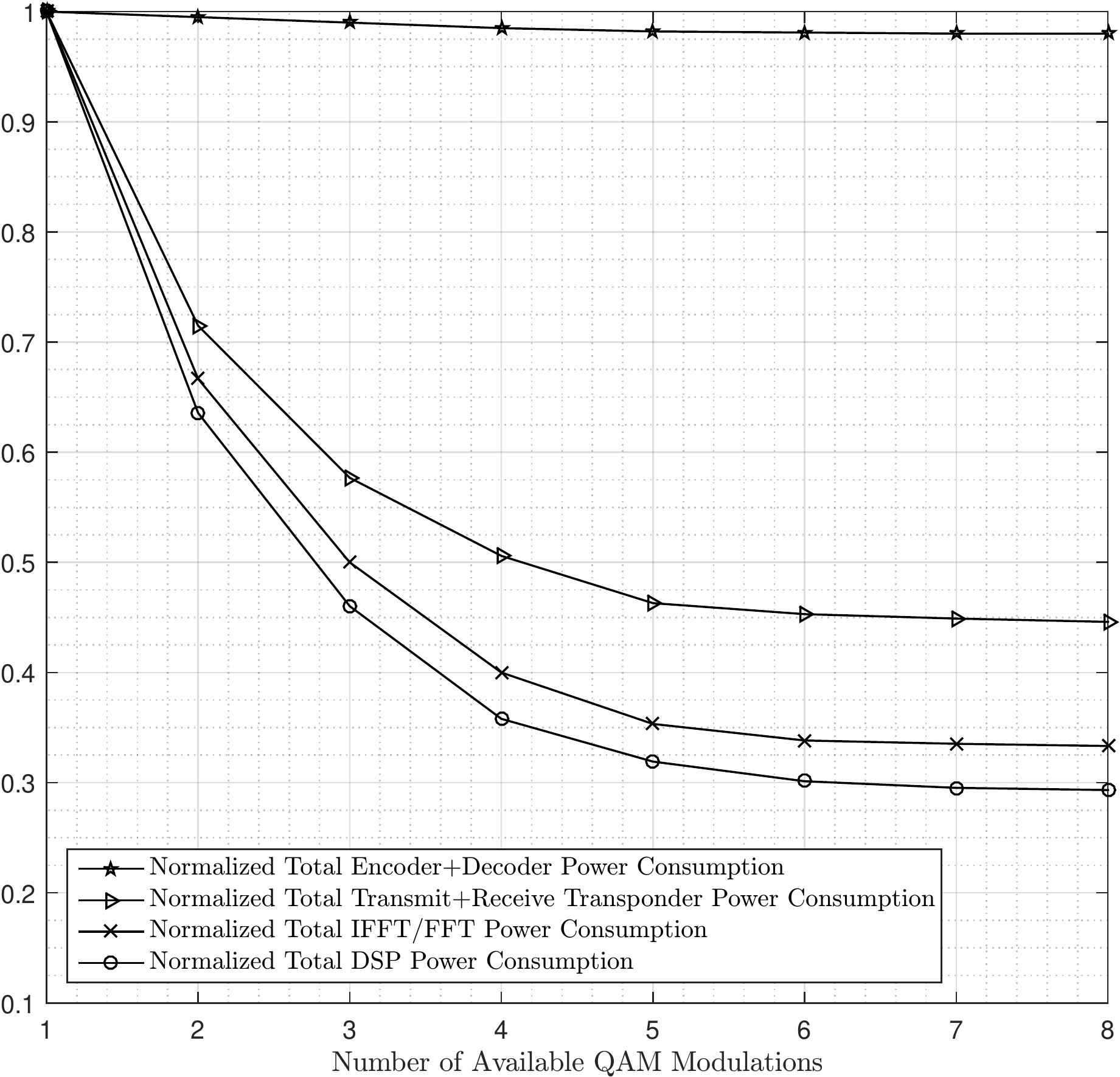}}
\center{\caption{\label{fig:levelAdapt} Normalized total power consumption of different network elements in terms of the number of available modulation formats.}}
\end{figure}

The effect of adaptive assignment of coding rate and modulation level on power consumption has been investigated in literature \cite{khodakarami2014flexible}. We also analyze this issue and evaluate the complexity of transponders. Fig. \ref{fig:levelAdapt} shows total power consumption of different network elements in terms of the number of available modulation formats. The power consumption values are normalized to their corresponding values for the scenario with two available modulation formats $c = 1, 2$. An interesting observation in Fig. \ref{fig:levelAdapt} is that increasing the number of modulation levels decreases the power consumption of various network elements with the DSP power consumption  showing the highest level of power reduction. Clearly, as the number of modulation levels increases, the amount of power consumption decreases but there is no considerable gain if the number of modulation levels be more than $6$. On the other hand, increasing the number of modulation levels in a transponder may have an important effect on the complexity of the receiver architecture and its signal processing requirements. Therefore, one should choose the number of modulation levels in a transponder to have a balance between complexity (CAPEX) and power consumption (OPEX).

\begin{figure}[t!] 
\center{\includegraphics[scale=0.5]{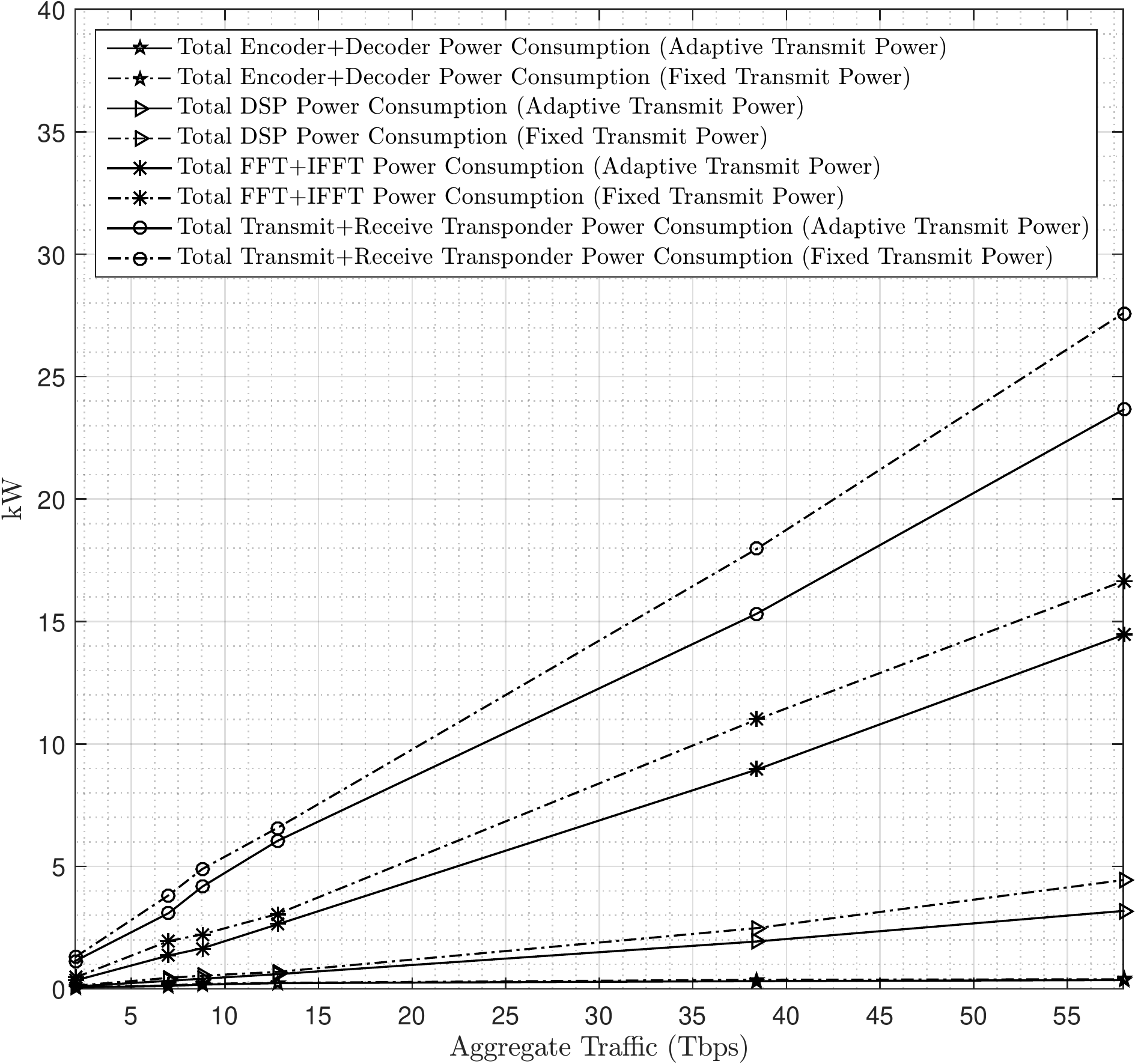}}
\center{\caption{\label{fig:powerAdapt} Total power consumption of different network elements in terms of aggregate traffic with and without adaptive transmit optical power assignment.}}
\end{figure}

The effect of adaptive assignment of coding rate and modulation level on power consumption has been studied in the literature. However, to the best of our knowledge, there is no comprehensive investigation for the effect of adaptive transmit optical power assignment on power consumption of different network elements. The total power consumption of different network elements in terms of aggregate traffic with and without adaptive transmit optical power assignment has been reported in Fig. \ref{fig:powerAdapt}. We have used the proposed expression of \cite{gao2012analytical} for fixed assignment of transmit optical power. Clearly, for all the elements, the total power consumption is approximately a linear function of aggregate traffic but the slope of the lines are lower when transmit optical powers are adaptively assigned. As an example, adaptive transmit optical power assignment improves total transponder power consumption by a factor of $8\%$ for aggregate traffic of $60$ Tbps. 

\section{Conclusion}\label{Sec_VII}
Energy-efficient resource allocation and quality of service provisioning are two fundamental problems of green elastic optical networks that are conventionally formulated using complex mixed-integer nonlinear optimization problems. To address the complexity of the problem and provide a fast-achievable near-optimum solution, we propose a two-stage resource allocation algorithm in which the main problem is decomposed into two inter-connected sub-problems named routing, grooming and ordering, and transponder parameter assignment. We introduce a heuristic approach for routing, grooming and ordering sub-problem that determines traffic routes, active transponders and grooming switch configuration to reduce amplifier power consumption and number of active transponders. To have an efficient transponder parameter assignment, we formulate a convex optimization problem in which joint optimization of power and spectrum resources constrained to physical and quality of service requirements is addressed. We use geometric convexification techniques to provide convex expressions for transponder power consumption and optical signal to noise ratio as a metric of quality of service. The performance of the proposed resource allocation procedure is validated through simulation results. We demonstrate the improved performance of our presented routing and grooming procedure in terms of network power consumption, transponder utilization ratio and traffic grooming ratio. Furthermore, numerical outcomes show that the proposed convex formulation for configuring the transponders yields the same results in a considerable lower run time compared to its mixed-integer nonlinear counterpart. The improved effect of the joint optimization of power and spectrum variables on power consumption of different network elements is also demonstrated by simulation. According to the results, adaptive assignment of transmit optical power considerably reduces total transponder power consumption. Furthermore, we analyze the inherent trade-off between CAPEX and OPEX as the capacity and/or design complexity of the transponders change.

\bibliographystyle{IEEEtran}
\bibliography{Reference}

\end{document}